# Model-Independent Comparison of Direct vs. Indirect Detection of Supersymmetric Dark Matter

Marc Kamionkowski[a,b,‡], Kim Griest[c,*], Gerard Jungman[d,†], and Bernard Sadoulet[e,⊙]

[a] School of Natural Sciences, Institute for Advanced Study, Princeton, NJ 08540
[b] Department of Physics, Columbia University, New York, NY 10027
[c] Department of Physics, University of California, San Diego, La Jolla, CA 92093
[d] Department of Physics, Syracuse University, Syracuse, NY 13244
[e] Center for Particle Astrophysics, University of California, Berkeley, CA 94720

ABSTRACT

We compare the rate for elastic scattering of neutralinos from various nuclei with the flux of upward muons induced by energetic neutrinos from neutralino annihilation in the Sun and Earth. We consider both scalar and axial-vector interactions of neutralinos with nuclei. We find that the event rate in a kg of germanium is roughly equivalent to that in a $10^5$- to $10^7$-$m^2$ muon detector for a neutralino with primarily scalar coupling to nuclei. For an axially coupled neutralino, the event rate in a 50-gram hydrogen detector is roughly the same as that in a 10- to 500-$m^2$ muon detector. Expected experimental backgrounds favor forthcoming elastic-scattering detectors for scalar couplings while the neutrino detectors have the advantage for axial-vector couplings.

1 December 1994

[‡] kamion@phys.columbia.edu
[*] kgriest@ucsd.edu
[†] jungman@npac.syr.edu
[⊙] sadoulet@lbl.gov

An ever-increasing body of evidence makes it feasible that the dark matter in our Galaxy is composed of weakly-interacting massive particles (WIMPs). The most promising candidate WIMP is probably the lightest supersymmetric particle, which in most cases is the neutralino, a linear combination of the superpartners of the photon, $Z$ boson, and Higgs bosons [1][2].

Several techniques are currently being pursued in an effort to discover such dark-matter particles. The first, direct detection (DD), seeks to observe the $\mathcal{O}(\text{keV})$ energy deposited in a low-background detector when a WIMP elastically scatters from a nucleus therein [3]. The second, indirect detection (ID), involves a search for energetic neutrinos produced by annihilation of WIMPs that have been captured in the center of the Sun and/or Earth [4].

Numerous calculations of event rates for both detection schemes for a variety of candidate WIMPs have been performed. Although supersymmetry (SUSY) is well-motivated and theoretically highly developed, there are many undetermined parameters, even in the minimal SUSY extension of the standard model (MSSM), so the results of any specific calculation often depend on a number of assumptions. Consequently, it is difficult to compare the constraints placed on WIMPs from DD experiments with those from ID experiments.

In this Letter, we provide a comparison of rates for DD and ID which is largely model independent. The rates for both techniques depend primarily on the coupling of WIMPs to nuclei. WIMPs couple both to the mass of a nucleus through a scalar interaction and to the spin through an axial-vector (spin) interaction. Here we focus only on WIMPs with either scalar or spin interactions; the results for a general WIMP should fall in between. In the end, we obtain the ratio of the rate per kg for elastic scattering from a variety of nuclei versus the flux of upward muons per square meter induced by energetic neutrinos from annihilation in the Sun and Earth.*

We account for the most significant model dependence quantitatively, and we discuss some residual model dependence which cannot be considered in a general fashion. The final results are to be taken as approximate and general results for a broad class of realistic WIMP candidates. A more precise comparison can, of course, be made for any specific model. It is also easy to imagine models with DD/ID ratios which fall far from our estimates. We have checked our results, however, by explicitly calculating and comparing DD and ID rates for thousands of allowed parameter choices in the MSSM [2].

This analysis should be helpful in assessing the relative effectiveness of the two methods of probing WIMPs with various couplings over a wide mass range, and for comparing various materials for low-background detectors. We caution that there are numerous experimental factors which we address only briefly that must be considered to properly weigh the relative sensitivities of the experiments to WIMPs. Although we have neutralinos in mind, the results for other WIMP candidates, such as heavy Dirac or Majorana neutrinos, should be similar.

We begin with particles with scalar interactions and note that the cross section for neutralino scattering from nucleus $i$ via a scalar interaction can be written

$$\sigma_i^{\text{sc}} = \frac{4 m_{\tilde\chi}^2 m_i^4}{\pi (m_{\tilde\chi} + m_i)^2} |\langle \mathcal{L}_{\text{sc}} \rangle|^2, \qquad (1)$$

where $m_{\tilde\chi}$ is the neutralino mass, $m_i$ is the nuclear mass, and $\langle \mathcal{L}_{\text{sc}} \rangle$ is the nucleon matrix element (scaled by the nuclear mass) of the effective Lagrangian for the scalar neutralino-nucleus interaction. The important thing to note is that all the information needed about any specific MSSM (e.g., the neutralino composition, the masses and couplings of all the superpartners, etc.) for the scalar neutralino-nucleus interaction is encoded in $\langle \mathcal{L}_{\text{sc}} \rangle$, and $\langle \mathcal{L}_{\text{sc}} \rangle$ is independent of the nuclear mass. To a great extent, both detection schemes provide constraints on $\langle \mathcal{L}_{\text{sc}} \rangle$. Moreover, theoretical uncertainties, such as the strange-quark scalar density in the nucleon, are absorbed in $\langle \mathcal{L}_{\text{sc}} \rangle$. These uncertainties affect both rates in the same way, so they do not affect the comparison between DD and ID.

Given the cross-section for elastic scattering of WIMPs from a nucleus, it is straightforward to compute the DD event rate. With Eq. (1), the rate for DD of scalar-coupled WIMPs is given by Eq. (18) in [6]. The event rate for scalar-coupled WIMPs in a detector composed of nuclei with mass number $A$, scaled by the rate in a $^{76}$Ge detector, is shown in Fig. 1 as a function of the WIMP mass.

---

\* A similar, though less comprehensive, comparison has been done by Rich and Tao [5].

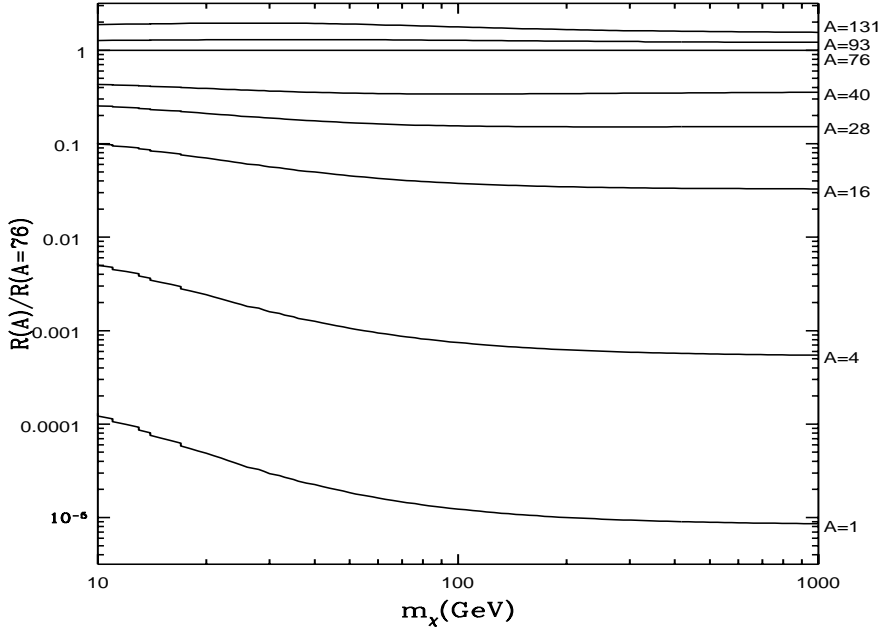

FIG. 1. Event rate (per kg of detector) for scalar-coupled WIMPs in a detector composed of nuclei with mass number $A$ scaled by the rate in a $^{76}$Ge detector as a function of WIMP mass $m_{\tilde{\chi}}$.

Energetic neutrinos from WIMP annihilation in the Sun or Earth are potentially detectable via observation of upward muons. The flux of such muons from WIMP annihilation in the Sun or Earth can be written

$$\Gamma = d \tanh^2(t/\tau)\, \rho_\chi^{0.3}\, f(m_{\tilde{\chi}})\xi(m_{\tilde{\chi}})(m_{\tilde{\chi}}/\text{GeV})^2 (\langle \mathcal{L}_{sc}\rangle /\text{GeV}^{-3})^2, \qquad (2)$$

where $d_\odot = 2.9 \times 10^8$ m$^{-2}$ yr$^{-1}$, and $d_\oplus = 1.5 \times 10^8$ m$^{-2}$ yr$^{-1}$, and it should be noted that the quantities $f(m_{\tilde{\chi}})$, $\xi(m_{\tilde{\chi}})$, and $t/\tau$ are different for annihilation in the Earth than they are for the Sun.

The capture rates in the Sun and Earth are $C_\odot = (2.1 \times 10^{37}\ \text{sec}^{-1})\, \rho_\chi^{0.3} f_\odot(m_{\tilde{\chi}})(\langle \mathcal{L}_{sc}\rangle /\text{GeV}^{-3})^2$ and $C_\oplus = (2.1 \times 10^{28}\ \text{sec}^{-1})\, \rho_\chi^{0.3} f_\oplus(m_{\tilde{\chi}})(\langle \mathcal{L}_{sc}\rangle /\text{GeV}^{-3})^2$.

The neutralino-mass dependence is described by the function

$$f(m_{\tilde{\chi}}) = \sum_i f_i \phi_i S_i(m_{\tilde{\chi}}) F_i(m_{\tilde{\chi}}) m_i^3 m_{\tilde{\chi}}/(m_{\tilde{\chi}} + m_i)^2, \qquad (3)$$

where the sum is over nuclei in the Sun or Earth, the quantities $f_i$ and $\phi_i$ are mass fractions and mean scaled potentials, and $S_i(m_{\tilde{\chi}})$ and $F_i(m_{\tilde{\chi}})$ describe resonance effects and form-factor suppression (see [2] and [7]).

The functions $\xi(m_{\tilde{\chi}})$ encode information about the neutrino spectra and the production of muons, and can be written,

$$\xi(m_{\tilde{\chi}}) = \sum_F B_F [3.47 \langle Nz^2 \rangle_{F,\nu}(m_{\tilde{\chi}}) + 2.08 \langle Nz^2 \rangle_{F,\bar{\nu}}(m_{\tilde{\chi}})], \qquad (4)$$

where the sum is over all annihilation channels $F$ available to the WIMP, and $B_F$ is the branching ratio for annihilation into $F$. The $\langle Nz^2 \rangle$ are scaled second moments of the neutrino (and antineutrino) energy distribution from final state $F$ for a given neutralino mass [8]. Neutrinos are absorbed in the Sun but not the Earth, so the $\langle Nz^2 \rangle$ are different for annihilation in the Sun than they are for annihilation in the Earth.

The most significant model dependence in our calculation comes in Eq. (4). There are many annihilation channels available to any specific neutralino candidate (and the number is larger for larger neutralino masses), and the $B_F$ may depend on a variety of couplings and particle masses. This leads to a range of values of $\xi(m_{\tilde{\chi}})$ for any given neutralino mass. However, the function $\xi(m_{\tilde{\chi}})$ will be bracketed above (below) by the value obtained from the annihilation channel which gives the largest (smallest) $\xi(m_{\tilde{\chi}})$. For neutralinos heavier than the top quark, the upper (lower) limit to $\xi(m_{\tilde{\chi}})$ comes from annihilation into top quarks (gauge bosons). If the WIMP is less massive than the $W$ boson, then the upper (lower) limit comes from annihilation into $\tau\bar{\tau}$ ($b\bar{b}$) pairs. If $m_W < m_{\tilde{\chi}} < m_t$, then the upper (lower) limit comes annihilation into gauge bosons ($\tau\bar{\tau}$ pairs).

The factor, $\tanh^2(t/\tau)$, in Eq. (2) describes the suppression of WIMP annihilation in the Sun or Earth relative to capture. Here, $t_\odot \simeq t_\oplus \simeq 4.5$ Gyr is the age of the solar system, and the $\tau$ are equilibration time scales given by $(t/\tau) =$ $r \left[\rho_\chi^{0.3} f(m_{\tilde\chi})(\sigma_A v)_{26}\right]^{1/2} (m_{\tilde\chi}/\text{ GeV})^{3/4}(\langle\mathcal{L}_{\text{sc}}\rangle/\text{ GeV}^{-3})$, where $(\sigma_A v)_{26}$ is the total annihilation cross section times relative velocity in the limit $v \to 0$ in units of $10^{-26}$ cm$^3$ sec$^{-1}$. Here, $r_\odot = 2.9 \times 10^7$ and $r_\oplus = 5.2 \times 10^4$. The analogous expression for the Earth is obtained by making the replacements $2.9 \times 10^7 \to 5.2 \times 10^4$ and $\odot \to \oplus$. If $\langle\mathcal{L}_{\text{sc}}\rangle$ is large, then $t/\tau \gg 1$, $\tanh(t/\tau) \simeq 1$, annihilation and capture are in equilibrium, and the signal is at full strength. In this case, $\Gamma \propto \langle\mathcal{L}_{\text{sc}}\rangle^2$, as is the DD rate. On the other hand, if $\langle\mathcal{L}_{\text{sc}}\rangle$ is small, then $t/\tau \ll 1$, annihilation has not had time to equilibrate with accretion and the neutrino signal is suppressed. In this case, $\Gamma \propto \langle\mathcal{L}_{\text{sc}}\rangle^4 (\sigma_A v)$.

To proceed, we make the simplest and most attractive assumption that if neutralinos exist, their abundance is suitable for accounting for a flat Universe. Then, $\Omega_\chi h^2 \simeq 0.25$ which fixes $(\sigma_A v)_{26} \simeq 1$. We then consider models which give neutrino fluxes in the range $10^{-2} \gtrsim \Gamma/(\text{m}^{-2}\text{ yr}^{-1}) \gtrsim 10^{-4}$ only. Models with larger fluxes would have been observed already, and the lower limit is roughly the sensitivity attainable with next-generation $\mathcal{O}(\text{ km}^2)$ detectors (accounting for the irreducible background of atmospheric neutrinos). We can then show that if $\Gamma(\sigma_A v)_{26}(m_{\tilde\chi}/\text{ GeV})^{-1/2} r^2 \gtrsim d\xi(m_{\tilde\chi})$, then the signal is at full strength. Taking $\Gamma_\odot > 10^{-4}$ m$^{-2}$ yr$^{-1}$, and $\xi(m_{\tilde\chi}) \lesssim 0.25$ (the maximum value of $\xi$ for any annihilation branch is 0.25), we find that the neutrino signal from the Sun is at full strength unless $m_{\tilde\chi} \gtrsim 10$ TeV. On the other hand, the signal from the Earth is potentially suppressed for any $m_{\tilde\chi} \gtrsim 10$ GeV.

The results for the comparison of the rate for elastic scattering from Ge with the flux of upward muons for scalar-coupled WIMPs are shown in Fig. 2 as a function of the WIMP mass. We consider the sum of muons from neutrinos from WIMP annihilation both in the Earth and in the Sun. The solid (dashed) curves are the ratios (including equilibration properly) for the upper (lower) limit for $\xi(m_{\tilde\chi})$. The upper (lower) pair of these curves are for WIMPs that give $\Gamma = 10^{-4} (10^{-2})$ m$^{-2}$ yr$^{-1}$. In the lower-limit curves, the neutrino signal from the Earth is essentially at full strength and is comparable to (for $m_{\tilde\chi} \gtrsim 80$ GeV) or greater than (for $m_{\tilde\chi} \lesssim 80$ GeV) the signal from the Sun. The model-dependent uncertainties are indicated by the range of values between the highest and lowest curves in this plot. The ratios for DD with other nuclei can be found using the scaling in Fig. 1.

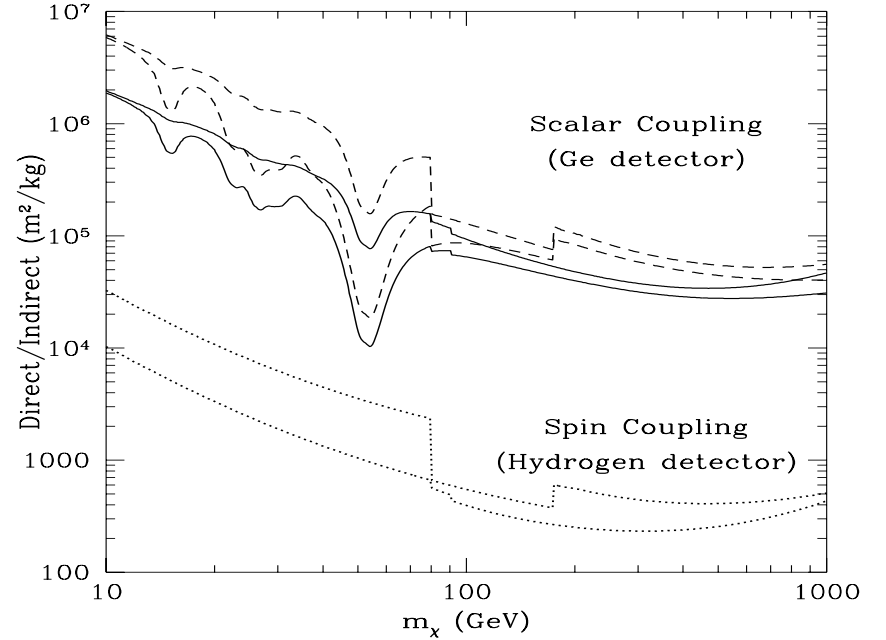

FIG. 2. Direct vs indirect detection of scalar- and spin-coupled WIMPs. For scalar-coupled WIMPs, we plot the ratio of the rate for elastic scattering from Ge in a laboratory detector to the flux of upward muons induced by neutrinos from annihilation in the Sun and Earth as a function of WIMP mass. The solid (dashed) curves are the ratios for the upper (lower) limit for $\xi(m_{\tilde\chi})$, the neutrino fluxes. The upper (lower) pair of these curves are for models that give $\Gamma = 10^{-4} (10^{-2})$ m$^{-2}$ yr$^{-1}$, and the model dependent uncertainties are indicated by the range of values between the highest and lowest curves. For scalar-coupled WIMPs, the ratios for detectors with different composition can be obtained using the scalings plotted in Fig. 1. For WIMPs with axial-vector couplings to nuclei, we plot ratios of the rate for elastic scattering from hydrogen in a laboratory detector to the flux of upward muons. The upper (lower) dotted curve is the ratio for the upper (lower) limit to the neutrino fluxes for spin-coupled WIMPs, and the model-dependent uncertainty is indicated by the range of values between these curves. In both cases, we neglect detector thresholds and backgrounds and assume efficiencies of order unity.

Our results show that for scalar couplings, the expected rate in 1 kg of Ge is roughly equivalent to that in $10^5 - 10^7$ m$^2$ of muon detector. This is the main conclusion of this Letter. To put it in perspective, we should briefly comment on the experimental situation. At least one DD experiment [9] using 1 kg Ge

5                                                                                                      6

and active background rejection techniques [10] should be operational within the coming year with a rejection factor of 99%. It will have a background of roughly 300 kg$^{-1}$ yr$^{-1}$ with an efficiency close to 100% for large enough masses.† Similarly the Dumand II [11] and AMANDA [12] collaborations are installing within similar time scales muon detectors with area of the order of $10^4$ m$^2$. The atmospheric-neutrino background will be roughly 300 events per year, coincidentally the same number of background events as for a kg of Ge. We expect therefore similar background rates for the coming generation of DD (1 kg of Ge) and ID ($10^4$ m$^2$) experiments. However, for scalar coupling, we have just shown that the rates are expected to be 10 to 1000 times larger for a 1-kg Ge experiment than for a $10^4$-m$^2$ neutrino detector, so the sensitivity of the DD experiment is much greater in this case.

We now turn to WIMPs with only a spin coupling. These WIMPs are captured in the Sun via scattering from hydrogen (H), but they are not captured in the Earth, and they may be detected directly through scattering only from nuclei with spin. The spin coupling to protons differs from that to neutrons, and the spin in heavier nuclei is generally carried at least in part by an unpaired neutron [13]. Furthermore, there may be a significant ambiguity in the relation between coupling to neutrons and protons due to uncertainties in the measured spin content of the nucleon [13][14]. Therefore, a model-independent comparison between ID and DD via scattering from an arbitrary nucleus cannot be made.

We can still, however, make a model-independent comparison of rates for ID with rates for DD in a detector made of H [15]. With Eqs. (10) and (25) in Ref. [7] and Eq. (18) in Ref. [6], we find,

$$[\text{Direct}(H)/\text{Indirect}] = 1.1 \times 10^5 m_{\tilde{\chi}}^{-2} [\xi(m_{\tilde{\chi}}) S_i(m_{\tilde{\chi}}, m_H)]^{-1} (\text{ kg}^{-1}/\text{ m}^{-2}). \quad (5)$$

---

† We use here the Ge experiment only for the purpose of illustration. A similar comparison can be made for the NaI experiments being constructed and for the sapphire and LiF cryogenic detectors being developed.

These results, for the allowed range of $\xi(m_{\tilde{\chi}})$, are shown in Fig. 2. Capture of spin-coupled WIMPs in the Sun occurs only via scattering from H. Therefore, there are no uncertainties from nuclear physics or nucleon spin structure that enter into this calculation. Moreover, if a WIMP has some scalar coupling as well, there will be additional capture in the Sun and Earth by scattering from heavier elements. Therefore, Eq. (5) gives a fairly robust upper bound to the DD/ID ratio for spin-coupled WIMPs. It appears possible to have in the coming years, about 50 g of H in a low pressure time-projection chamber [15] which should have negligible background. The corresponding limit on the rate would be after 1 year, about 50 kg$^{-1}$ yr$^{-1}$ which is therefore equivalent to a muon-flux limit of 0.005 (at low mass) to 0.25 (at large masses) m$^{-2}$ yr$^{-1}$. This has already or will soon be reached by the current neutrino experiments. The indirect method has therefore a clear advantage in this case.

Let us assume for the moment that the spin coupling to neutrons is the same as that for protons. There is no strong reason to believe this is true in a particular model, but this will allow us to scale approximately the dependence of the rate on the nucleus. Then, the scaling of the rate for DD of spin-coupled WIMPs for other nuclei with mass $m_i$ is roughly $f_{\text{spin}} \eta_c(m_{\tilde{\chi}}, m_i) m_i / (m_{\tilde{\chi}} + m_i)^2$, where $f_{\text{spin}}$ is the mass fraction of the given isotope in the detector. The function $\eta_C$ is the DD form-factor suppression given in Eq. (19) in [6], although we caution that the functional form for the spin interaction may differ somewhat [16]. The Ge experiment quoted above intends to use 500 g of isotopically pure $^{73}$Ge. Assuming the same backgrounds as before the Ge experiment appears roughly equivalent to a $10^4$-m$^2$ indirect experiment. However, this is highly dependent on details of the spin content of the nucleon, and usual estimates [13] lead to a substantial DD rate reduction compared to this.

To conclude, we have found that for scalar-coupled WIMPs, the event rate for DD in a kg of Ge is roughly the same as the rate in a neutrino telescope of area $10^5$ to $10^7$ m$^2$ for equal exposure times. For spin-coupled WIMPs, the DD event rates in a 50-g H detector is roughly equivalent to that in a 10- to 500-m$^2$ detector. We have also shown how these results can be scaled to DD rates for other target nuclei, although the scaling may be quite model dependent for spin-coupled WIMPs.

Taking into account expected experimental backgrounds, the forthcoming 1-kg Ge experiment with active background rejection should have an advantage over the $10^4$-m$^2$ detectors under construction in case of dominant scalar interactions. However, in the case of dominant spin interactions, the above neutrino experiments should have the advantage: the isotopically enriched $^{73}$Ge experiment is expected to have only comparable sensitivity at best to that of the above neutrino detectors, and the sensitivities of 50-g H detectors will be even lower.

In realistic SUSY models, DD via scalar interactions tends to dominate DD via spin interactions by factors of thousands over much of parameter space. On the other hand, the spin interaction plays a much more important role in capture of WIMPs in the Sun. Therefore, for a general WIMP the ratio of DD versus ID rates will fall somewhere between the results for a purely scalar-coupled and a purely spin-coupled WIMP. However, by explicit evaluation of the DD/ID ratio in several thousand realistic SUSY models, we find that in most regions of parameter space, the DD/ID ratio seems to be well described by the curves for purely scalar-coupled WIMPs shown in Fig. 2. This does not mean that there are not regions of SUSY parameter space where spin coupling dominates, but it is likely that a generic neutralino will be primarily scalar coupled, and so the upper curves in Fig. 2 will apply. It should also be noted that there may be other (perhaps non-SUSY) WIMP candidates (such as Majorana neutrinos) that have only spin couplings.

It is a pleasure to thank F. Halzen, C. Tao, and J. Engel for useful discussions. B.S. thanks M. Runyan for assistance. M.K. acknowledges the hospitality of the CERN Theory Group where part of this work was completed. This work was supported in part (B.S. and K.G.) by the Center for Particle Astrophysics, a NSF Science and Technology Center operated by the U. of California under Cooperative Agreement No. ADT-8809616. M.K. was supported by the W. M. Keck Foundation at the I.A.S. and by the D.O.E. at Columbia under contract DEFG02-92-ER 40699. G.J. was supported by the D.O.E. under contract DEFG02-85-ER 40231. K.G. was supported in part by a D.O.E. OJI award and the Alfred P. Sloan Foundation.